\documentclass{ifacconf}

\usepackage{graphicx}      
\usepackage{natbib}        
\usepackage{float} 
\begin{document}
\begin{frontmatter}

\title{Autonomous Industrial Control using an Agentic Framework with Large Language Models} 

\thanks[footnoteinfo]{The authors gratefully acknowledge the financial support provided by the Department of Chemical Engineering at Imperial College London for this work}

\author{Javal Vyas},
\author{Mehmet Mercangöz}

\address{Autonomous Industrial Systems Lab, Imperial College London, Imperial College Rd, South Kensington Campus, London, SW7 2AZ, United Kingdom}

\begin{abstract}                
As chemical plants evolve towards full autonomy, the need for effective fault handling and control in dynamic, unpredictable environments becomes increasingly critical. This paper proposes an innovative approach to industrial automation, introducing validation and reprompting architectures utilizing large language model (LLM)-based autonomous control agents. The proposed agentic system—comprising of operator, validator, and reprompter agents—enables autonomous management of control tasks, adapting to unforeseen disturbances without human intervention. By utilizing validation and reprompting architectures, the framework allows agents to recover from errors and continuously improve decision-making in real-time industrial scenarios. We hypothesize that this mechanism will enhance performance and reliability across a variety of LLMs, offering a path toward fully autonomous systems capable of handling unexpected challenges, paving the way for robust, adaptive control in complex industrial environments. To demonstrate the concept's effectiveness, we created a simple case study involving a temperature control experiment embedded on a microcontroller device, validating the proposed approach.
\end{abstract}


\end{frontmatter}
\section{Introduction}
Chemical plants are moving towards autonomous operations. Especially for routine operations that follow well-defined procedures, autonomous operation is considered technically feasible with currently available technologies (\cite{borghesan2022unmanned}). However, a significant challenge in developing autonomous control systems is the need to account for long-tail events, which are rare, unpredictable occurrences that fall outside of the scope of typical operational scenarios. In industrial contexts, these long-tail events can range from unexpected equipment failures to highly unusual process disturbances. Traditional automation approaches struggle to handle such events, as they rely heavily on predefined rules and algorithms, rendering them overly rigid and poorly adapted to situations that deviate from expected patterns. Solutions leveraging machine learning models have made some progress in handling known unknowns such as known disturbances or possible plant-model mismatch but they tend to fail in handling anomalies. This is primarily because these models are trained on majority-class data, as anomaly data is scarce or available in too few samples. As a result, these solutions struggle to detect and react to anomalies in real-time, particularly in scenarios involving unknown unknowns—unforeseen disturbances that the system was not designed to handle.

Currently, human operators play a key role in managing the type of  unknown unknowns discussed previously. Leveraging their reasoning abilities and domain knowledge human operators can dynamically assess a situation and adjust their actions based on real-time feedback. The overarching goal of this work is to bridge these reasoning and knowledge use abilities to autonomous systems using generative machine learning models as intelligent control agents. We particularly focus on the use of Large Language Models (LLMs) for this purpose.

LLMs, with their extensive knowledge bases and reasoning capabilities, represent a promising avenue for developing intelligent control agents capable of autonomously analyzing incoming data, diagnosing anomalies, and making informed control decisions in a zero-shot manner- making inferences and offering solutions to scenarios they have not explicitly encountered in training \citep{Pantelides2024}. The challenge is transitioning to a fully automated system that can evaluate responses and adjust actions independently. To address this, we propose a reprompting architecture that empowers LLMs to function as autonomous control agents. This architecture enables agents to validate their actions against a digital twin, implementing them in the physical system if they pass validation; if not, the agent is prompted to revise its approach. This iterative process significantly enhances decision-making capabilities and improves system performance in real-time.

\section{Related Work}
\subsection{Evolution of Autonomous Systems for Industrial Control}
Autonomous systems have been defined in various ways across the literature, with some emphasizing their capability to solve tasks independently of specific programming instructions (\cite{AutonomyDef}, \cite{abbass2018foundations}) and others noting their ability to achieve goals without step-by-step guidance (\cite{beer2014toward}, \cite{watson2005autonomous}). Another perspective highlights autonomy as the capacity to make decisions under incomplete information (\cite{abbass2018foundations}, \cite{aniculaesei2018towards}). These definitions underscore the growing role of artificial intelligence (AI) in industrial control systems, positioning Autonomous Industrial Systems as a key area within Industrial AI, intersecting with fields like Machine Learning, Natural Language Processing, and Robotics (\cite{IndustrialAI}).

Initial approaches in agent based systems utilized rule-based agents for tasks such as intrusion detection (\cite{Rule_based1}) or decision support (\cite{Rule_based2}). Despite some success, rule-based agents are inherently limited to predefined situations, making them less adaptable to novel scenarios (\cite{Human&AI}). This constraint led researchers to explore machine learning and deep learning (DL) agents, which can adapt based on data. For instance, DL-based multi-agent systems have shown effectiveness in intrusion detection (\cite{DL_1}), yet the complexities of data collection and validation in distributed environments create substantial challenges in many industrial applications (\cite{DL_2}).

To address these limitations, researchers turned to reinforcement learning (RL) agents. RL agents, while highly effective for specialized tasks, face challenges in sample efficiency, generalizability, and lengthy training times (\cite{ExploringAgents}). Despite their effectiveness in specific applications like process control in crystallization (\cite{RL_1}) and inventory management (\cite{RL_2}), RL-based approaches often require well-defined problem settings and reward functions, which can limit their scalability in complex, dynamic environments (\cite{RL_3}) and may not be well suited for handling anomalous conditions.

\subsection{Large Language Models (LLMs) in Industrial Control}
Recently, large language models (LLMs) have emerged as a promising tool in agent based systems due to their adaptability and generalization abilities. LLMs have made significant inroads in chemical engineering, such as predicting material properties (\cite{LLMatDesign}, \cite{GPTMolBERTa}) and process decision-making (\cite{CanLi}, \cite{PILOT}). LLMs have also been employed for tasks like fault detection and flowsheet generation, where they assist in complex problem-solving by completing, correcting, or even generating flowsheets autonomously (\cite{FlowsheetCompletion}, \cite{FlowsheetGeneration}). 

LLMs have also been considered for industrial control as well. \cite{HVACLLM} proposed a framework to control the HVAC system in a building using an LLM. Researchers used the historical demonstrations along with the prompt to the LLM performance in controlling the HVAC system in the building. They demonstrated that an LLM performs equivalent or surpasses the RL performance. Researchers have also used LLMs for the modular production and control of autonomous industrial systems, where the LLMs are connetced to the digital twins and LLMs adapt with the interactions with the digital twin for a specific task (\cite{GPT4IAS}). \cite{LLM4IAS} propose a framework to achieve an end to end industrial automation system. Their framework
supplies LLMs with real-time events on different context
semantic levels, allowing them to interpret the information,
generate production plans, and control operations on the
automation system. In this work, the researchers propose to use a digital-twin of the industrial system for generating context for the automation agents but they do not consider a validation or reprompting scheme or the generation of any kind of feedback or critique for the LLM actions.

\subsection{Prompting Strategies for Enhanced Control}
Prompting strategies have become central to improving LLMs’ decision-making abilities in complex tasks. Among these, the Chain of Thought (CoT) approach prompts LLMs to break down tasks into intermediate reasoning steps before producing a final response (\cite{COT}). Building on this, the Tree of Thought (ToT) approach expands on CoT by allowing LLMs to explore multiple paths in their reasoning (\cite{TreeOfThought}), and the Graph of Thought (GoT) consolidates LLM reasoning paths to enhance task completion accuracy (\cite{GraphOfThought}).

Other prompting frameworks adapt feedback-driven approaches to improve LLM behavior in agent-based settings. REACT (\cite{React}) enables agents to process thoughts before taking actions, while REFLECTION (\cite{REFLEXION}) allows agents to interact with the system, reflect on actions, and store the interaction history for iterative learning. Related approaches, including self-refine (\cite{self-refine}), RCI (\cite{RCI}), and self-debugging (\cite{self-debugging}), utilize feedback for error-correction and optimization in domain-specific tasks. For iterative tasks, \cite{RePrompt} introduce a “gradient descent” style reprompting method to refine prompts based on interaction history, while \cite{reprompting} use automatic reprompting with CoT to improve task accuracy.

This progression toward LLMs highlights their potential to address the limitations of earlier methods, providing a versatile approach that combines interpretability and robustness for industrial automation. 

\begin{figure*}[t] 
\begin{center}
\includegraphics[width=0.85\textwidth]{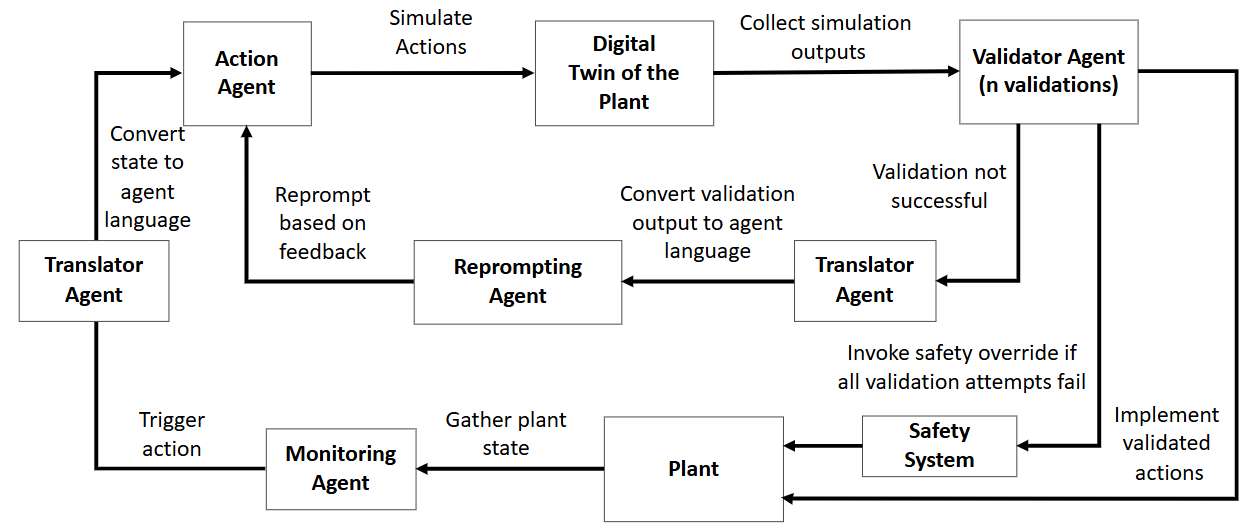} 
\caption{Schematic of an agentic framework for monitoring and controlling a process plant during anomalous conditions} 
\label{fig:Framework Schematic}
\end{center}
\end{figure*}

\subsection{Contribution}
In this work, we introduce the concept of using reprompting architectures for industrial control, where LLMs operate autonomously in complex process environments. The idea here is to have an agent based system which can carry out tasks autonomously. The engine for the agents in the system would be a large language model (LLM). We hypothesize that with the reprompting architecture, the performance of the system would improve. LLMs which are prone to hallucinations as their inherent model characteristic may result in erroneous action, which can be hazardous in safety critical systems. Thus, having another agent which acts as a critique in navigating to an safe-to-optimal response would decrease the likelihood of the error rates. More specifically we introduce validation agents utilizing a simulation capability e.g. using a digital-twin to check the utility of the actions generated by the LLM agents and use a reprompting agent to provide feedback to the actor agent for improving the action in case the previously suggested action does not pass the validation check. 

To illustrate the potential of this approach, we present a case study focused on temperature control using a physical micro-controller. We argue that this methodology aligns with the growing trend towards adaptive, fully autonomous systems and establishes a new pathway for intelligent industrial automation.

The following sections delve deeper into the components of the proposed framework. Section 3 provides an overview of the framework and its components. In section 4 we present the temperature control case study and its architecture. Section 5 discusses the results of the case study and its findings. Finally, section 6 we touch upon the future work.  

\section{Methodology}
The proposed framework introduces a modular and adaptive LLM-based multi-agent system, with a focus on programmatically leveraging a Reprompting step via a Reprompter Agent to guide an Actor Agent toward safe and effective solutions. Each agent is assigned a specific role, equipped with tools, and tasked with distinct actions that contribute to the overarching system objectives. This section outlines the framework's role in enhancing system reliability and responsiveness through a coordinated agent-based approach.
\subsection{Framework Overview}
The core of this framework is built around four principal agents—the Monitoring Agent, Actor Agent, Validator Agent, and Reprompter Agent—that interact with a simulated digital twin environment (see e.g. Fig 1). This digital twin serves as a proxy for the physical system, enabling safe validation of actions and structured feedback loops before passing actions to the physical plant.

\begin{itemize} \item \textbf{Monitor Agent:} The Monitoring Agent gathers the state from the plant and can act as a versatile agent. The Monitoring agent can be used for both continuous control or anomaly detection. In case of continuous control, it would keep the track of the performance of the system and would allow for a planned action in a continuous manner. While in case of anomaly detection, the Monitoring agent would only trigger the subsequent agents if it detects the anomaly. 

\item \textbf{Actor Agent:} The Actor Agent initiates actions aimed at achieving control objectives, such as modifying parameters or toggling operational states. It operates based on predefined goals, and once it formulates an action, the Actor Agent passes this decision to the digital twin. This simulation evaluates the potential effects of the action, minimizing the risk of unsafe interventions on the physical system.

\item \textbf{Digital Twin Simulation:} The digital twin emulates the behavior of the physical system in response to the Actor Agent’s actions, enabling real-time assessment in a no-risk environment. This simulated feedback captures anticipated system responses, allowing agents to test actions safely before deployment.

\item \textbf{Validator Agent:} Following the simulation, the Validator Agent assesses the Actor Agent’s proposed action based on safety and operational criteria. If the action meets these criteria, it is ready for physical deployment. However, if it is deemed unsafe or suboptimal, the Validator Agent flags the action, prompting the Reprompter Agent to intervene for a predefined iterations after which, if the actions are unsafe, the safety system would override the actions. 

\item \textbf{Reprompter Agent:} The Reprompter Agent is a pivotal component in ensuring system safety and refinement. When an action fails validation, the Reprompter Agent collaborates with the Actor Agent to adjust the initial decision. Using alternative prompts generated by processing the digital-twin outputs, the Reprompter Agent conditions the Action Agent until it aligns with the Validator Agent’s criteria. This process forms a feedback loop in which each iteration is tested in the digital twin and validated again, ensuring the action is both safe and optimized. The loop persists until the action either satisfies validation standards or reaches a predefined limit on iterations, safeguarding stability in the control process. \end{itemize}

This structured interaction between agents, anchored by the Reprompter Agent’s corrective capabilities, enables the system to autonomously navigate complex control environments. By leveraging programmatic refinement, the Reprompter Agent helps the Actor Agent reach safe and effective solutions, ensuring robust and adaptive control in dynamic industrial settings.


\subsection{Components of framework}
\begin{itemize}
    \item \textbf{Agents:} Each agent functions as a specialized LLM-driven entity with a distinct role in the feedback loop, contributing to adaptive control: 
    \begin{itemize}
        \item \textbf{Role:} Defines the purpose of each agent.
        \item \textbf{Goal:} Provides clarity on what each agent should achieve.
        \item \textbf{LLM:} Serves as the core reasoning engine, enabling agents to analyze, evaluate, and adapt.
        \item \textbf{Tools:} Specify tools which the agent would have access to in the decision making process. 
    \end{itemize}
    \item \textbf{Tools:} These serve as specialized utility functions that support agents during decision-making. Tools enable agents to handle tasks that are beyond the core capabilities of an LLM, such as performing complex calculations or accessing specific lookup tables. By supplementing the LLM’s reasoning with precise computational and data-access functions, the tools enhance the agents' ability to make informed, accurate decisions.

    \item \textbf{Tasks:} These are targeted assignments given to each agent, ranging from concise directives to detailed instructions that guide the LLM in executing specific actions. Tasks are carefully assigned to agents equipped with the necessary expertise or context, ensuring the agent’s background aligns with the requirements of the task. Each task description provides clear guidance to optimize agent performance and streamline the overall decision-making process.
\end{itemize}

In summary, this methodology outlines a structured, iterative framework designed to leverage the capabilities of Large Language Model (LLM)-based agents in autonomous industrial control. Each component, from specialized agents to supporting tools and defined tasks, works in concert to ensure safe, adaptive, and effective control actions within a digital twin environment. The introduction of a Reprompter Agent strengthens the system’s resilience, facilitating a feedback-driven refinement process that iteratively adjusts actions until they meet safety and efficacy standards.

To demonstrate the practical application of this framework, we present a case study in temperature regulation. This case study illustrates the roles and interactions of the Actor, Validator, and Reprompter agents in real-world scenarios, showcasing the framework’s capability to autonomously navigate complex control challenges.

\section{Case Study}
This case study demonstrates the application of the proposed LLM-based multi-agent framework to autonomously control a physical Arduino microcontroller known as TCLab (\cite{TCLab}). The setup aims to manage heater operations based on specific temperature thresholds: heaters are turned off when the temperature exceeds 27°C and turned on when it falls below 25°C. This creates a cyclical oscillation within these thresholds, with the control sequence monitored over a 40-minute period. The goal of this case study is to assess how effectively the proposed multi-agent framework improves via the use of re-prompting for autonomous control under real-world conditions.

\textbf{Structure of the Case Study}
The case study employs a three-agent structure (Fig 2), each with distinct roles to facilitate intelligent decision-making and control processes:
\begin{figure}
\begin{center}
\includegraphics[width=8.4cm]{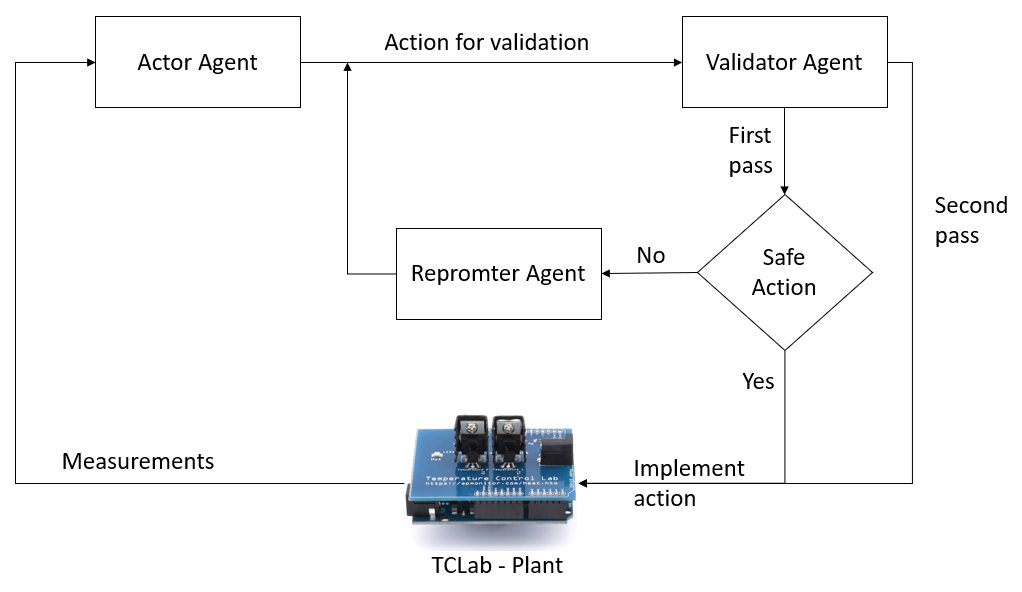}    
\caption{Case Study Schematic} 
\label{fig:Case Study Schematic}
\end{center}
\end{figure}
\begin{itemize}
    \item \textbf{Operator Agent:} The Operator agent initiates an action based on real-time temperature readings. It determines when to activate or deactivate the heater based on the predefined thresholds(\ref{fig:Task description}). By leveraging its role-specific prompts (\ref{fig:Agent Description}), the Operator Agent interprets data and issues commands to maintain the target temperature range.

    \item \textbf{Validator Agent:} The Validator Agent assesses the actions proposed by the Operator Agent. It verifies whether the action aligns with the control logic—specifically, maintaining temperatures within the desired range. If the proposed action does not meet the criteria, the Validator flags it for reevaluation, preventing potentially unsafe or incorrect responses from being implemented.

    \item \textbf{Reprompter Agent:} Upon a validation failure, the Reprompter Agent is activated to analyze and refine the action suggested by the Operator Agent. The Reprompter Agent recalibrates the initial action to ensure it aligns with the system's predefined requirements. The refined action undergoes a secondary validation before being deployed.
\end{itemize}

\begin{figure}[t]
\begin{center}
\includegraphics[width=8.4cm]{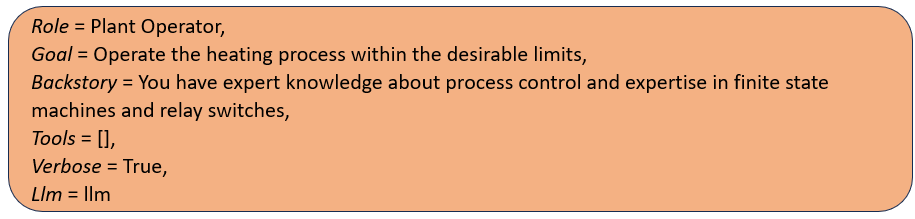}    
\caption{Operator Agent Description} 
\label{fig:Agent Description}
\end{center}
\end{figure}

\begin{figure}
\begin{center}
\includegraphics[width=8.4cm]{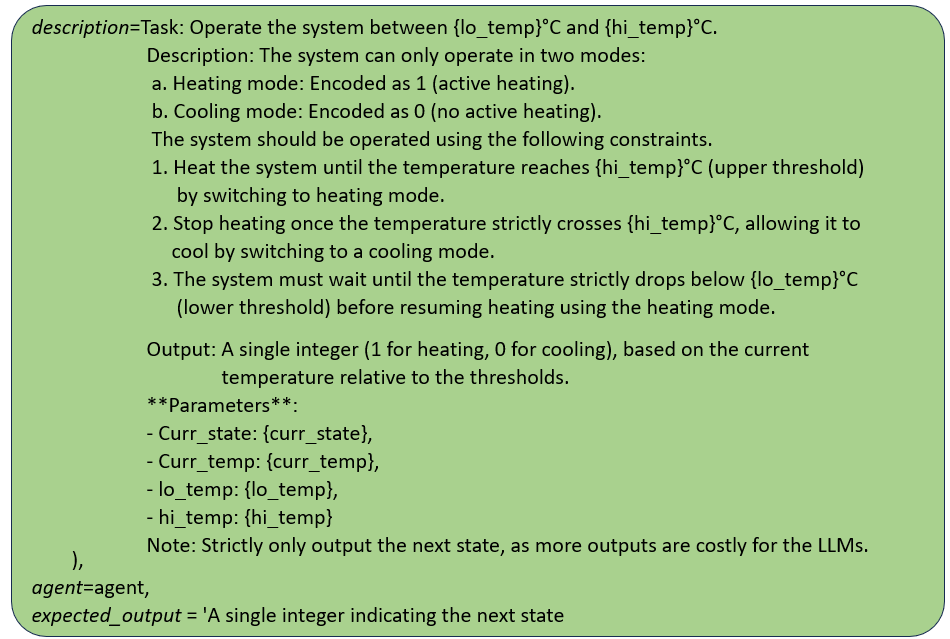}    
\caption{Operater Agent Task Description} 
\label{fig:Task description}
\end{center}
\end{figure}

While the framework is generally designed to work with a digital twin model, the simplicity of this control task allows us to embed it into the Validator Agent in this case study. This approach demonstrates the framework’s adaptability to different control tasks, highlighting each agent’s contribution to maintaining stable, autonomous control. Future work will integrate a digital twin, particularly for complex, safety-critical scenarios like fault detection, enhancing the framework’s robustness for advanced industrial control. This case study underscores the potential of this multi-agent configuration to autonomously manage and correct actions in real-world applications.

\section{Results}
The case study framework was realized using CrewAI mutli-agent platform \citep{crewai2024}. The framework was createdand executed locally while the agents' engine utilized LLMs from OpenAI suite \citep{openai2024} accessed in the cloud via the OpenAI APIs for the correspondding LLMs . The communication between TCLab and the framework was achieved via a Python based wrapper. The performance of the proposed framework, leveraging OpenAI's large language models (LLMs) suite as control agents, was evaluated within a temperature regulation case study. This evaluation centered on the agents’ accuracy in executing control actions and their control performance. We measured accuracy across two settings—initial pass accuracy and accuracy post-reprompting—to analyze the models’ ability to correct missteps autonomously.

\begin{table}[hb]
\begin{center}
\caption{Accuracy Performance of Language Models in the proposed framework}\label{table:performance}
\tiny
\begin{tabular}{@{}|c|c|c|c|c|@{}}\hline
 Metric & GPT3.5 & GPT4omini & GPT4o & GPT4\\\hline
Accuracy- first pass (\%) & 60.04 & 72.49 & 99.63 & 93.75\\
Accuracy - reprompts (\%)& 85.34 & 89.97 & 99.81 & 96.09\\
Samples & 423 & 394 & 554 & 128\\
Passes & 254 & 253& 552 & 120\\
Fails & 169 & 61 & 2 & 8\\
Pass after reprompts & 107 & 96& 1 & 3\\\hline
\end{tabular}
\end{center}
\end{table}

\begin{figure}[h]
\begin{center}
\includegraphics[width=8.4cm]{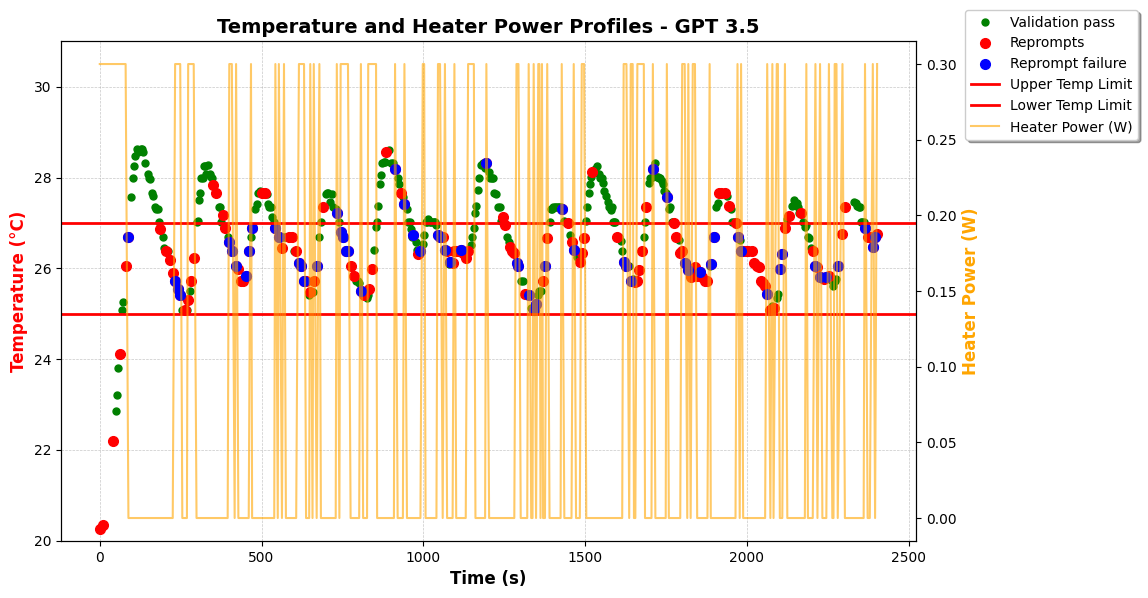}    
\caption{Temperature Profile for GPT 3.5} 
\label{fig:Temp gpt35}
\end{center}
\end{figure}

\begin{figure}[h]
\begin{center}
\includegraphics[width=8.4cm]{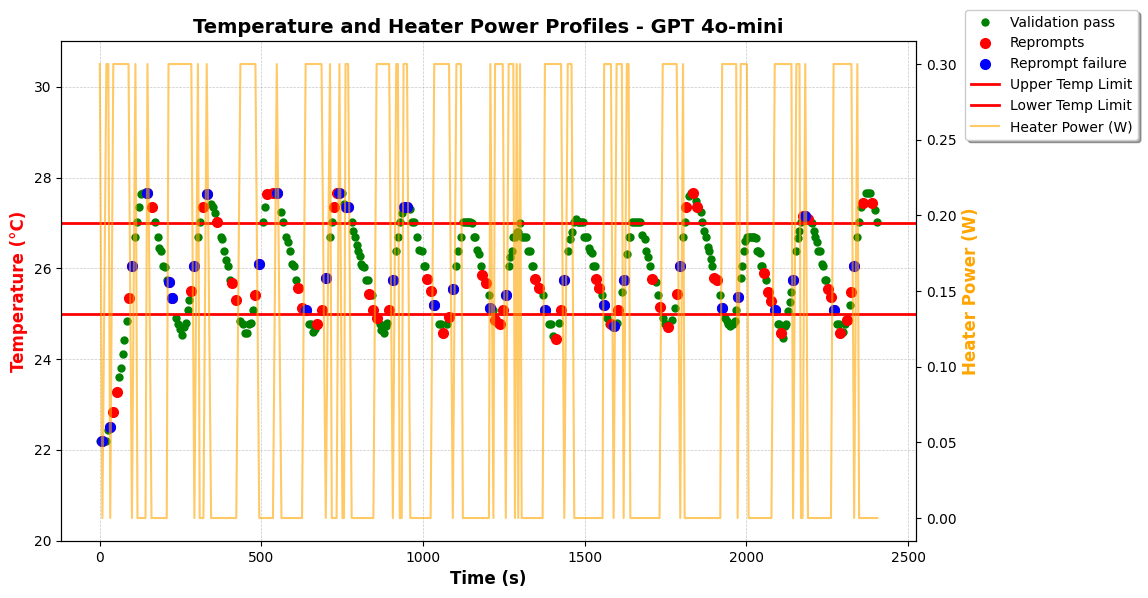}    
\caption{Temperature Profile for GPT 4o-mini} 
\label{fig:Temp gpt 4o-mini}
\end{center}
\end{figure}

\begin{figure}[h]
\begin{center}
\includegraphics[width=8.4cm]{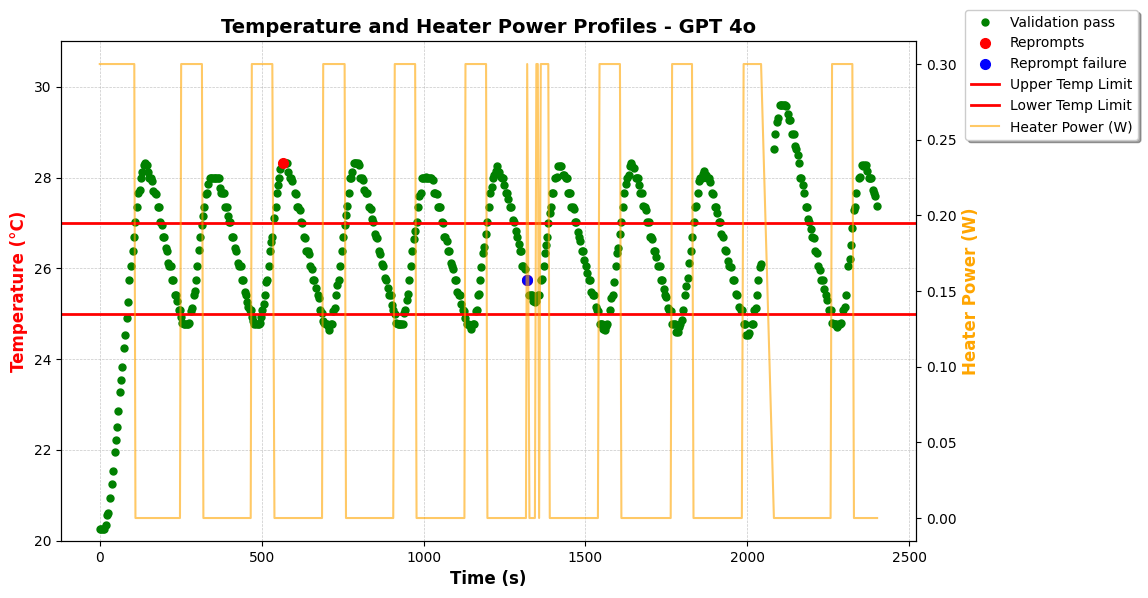}    
 \caption{Temperature Profile for GPT 4o} 
\label{fig:Temp gpt 4o}
\end{center}
\end{figure}

\begin{figure}[h]
\begin{center}
\includegraphics[width=8.4cm]{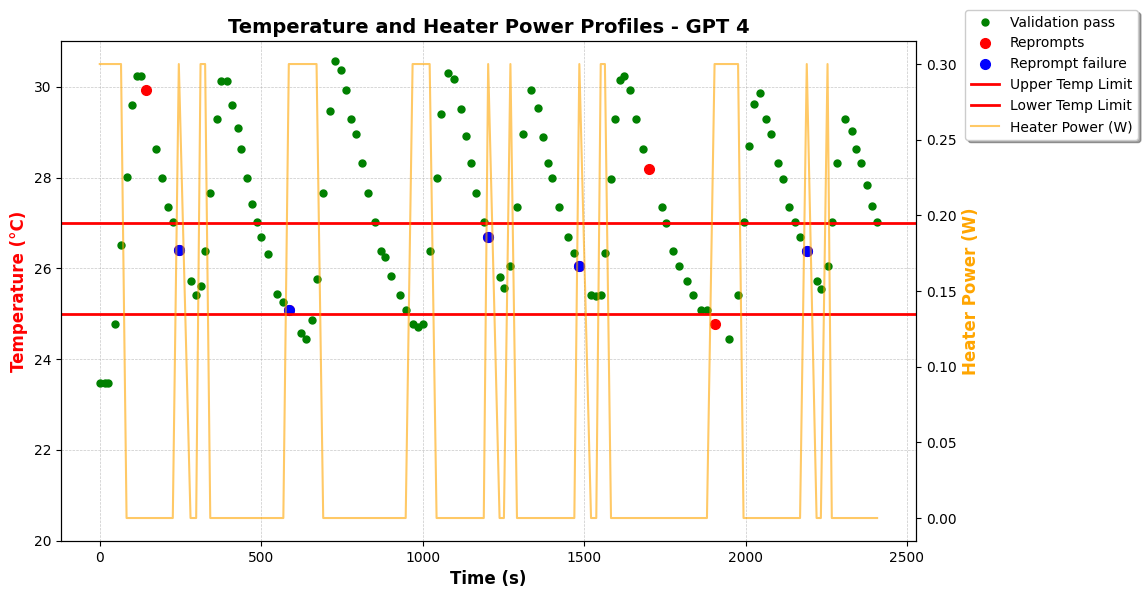}    
\caption{Temperature Profile for GPT 4} 
\label{fig:Temp gpt 4}
\end{center}
\end{figure}

The results in Table \ref{table:performance} show a two fold story, one is about the sampling rates and other about the accuracy. Here the sampling rates for GPT 4o is the highest while GPT 4 has the lowest sampling rate. This is a result of the inference time of these models, thus impact the sampling rates. This although of lesser importance for this application, indicates that autonomous systems with LLM based agents may not be suitable for a system that needs to have 
fast dynamics. 

In terms of accuracy, GPT 4o outperforms other OpenAI models, with GPT 4 following as the second-best performer. Notably, when reprompting is applied, the system's performance increases across all models, with the most significant improvement observed in GPT 3.5 rising from 60.4\% to 85.34\%. These results highlight the potential of reprompting architectures to enhance model performance significantly, enabling even less capable models to approach the accuracy of more advanced ones.  

The control performance of these models were evaluated using the average temperate deviation from the midpoint of the temperature range. Table \ref{table:control performance} shows that GPT 4o mini has the best control performance where the overshoots and undershoots are minimal, whereas while GPT 4 being highly accurate performs the worst in control performance amongst the OpenAI LLM suite. This is attributed to the inference time of the model. For GPT 4 the inference time was high resulting in previous action being implemented for an extended period of time. Since the system does not have cooling, even when the heaters are switched off,  residual heat continues to dissipate, raising the temperature further. Thus, it is important to note that LLM inference times do influence the control performance of the system.

\begin{table}[hb]
\begin{center}
\caption{Control Performance of Language Models in the proposed framework}\label{table:control performance}
\tiny
\begin{tabular}{@{}|c|c|c|c|c|@{}}\hline
 Metric & GPT3.5 & GPT4omini & GPT4o & GPT4\\\hline
Average Deviation & 0.832 & 0.077 & 0.582 & 1.469\\
Time above 27C (s)& 949 & 499.10 & 887.70 & 1163.09\\
Time below 25C (s) & 0 & 432.30 & 285.87 & 173.40\\
Time outside range (s) & 949.24 & 931.40& 1173.58 & 1336.50
\\\hline
\end{tabular}
\end{center}
\end{table}



These results confirm that the proposed framework effectively utilizes LLMs as control agents, and the reprompting mechanism significantly enhances accuracy and reliability, especially for models with initially lower performance.

\section{Conclusion}
In conclusion, this paper highlights the promising potential of large language models (LLMs) as autonomous control agents in industrial applications. The proposed framework, enhanced by a reprompting architecture, demonstrates a significant capability for agents to autonomously correct their actions, leading to improved reliability and accuracy in control tasks. Our results indicate that even earlier models like GPT 3.5-turbo can achieve substantial performance gains through reprompting, with accuracy improving from 60.04\% to 85.34\%. More advanced models, such as GPT 4o, reached near-perfect accuracy exceeding 99\%, showcasing the framework's effectiveness in harnessing LLMs for control tasks with proposed framework.

These findings validate the viability of LLM-based systems in autonomous industrial control, where rapid and precise decision-making is essential. While this case study focused on a relatively straightforward task, the adaptability of the framework positions it well for application in more complex control scenarios. Future work may explore its deployment in fault handling and digital twin environments, where real-time decision-making is critical in dynamic and unpredictable settings. Overall, this research supports the integration of LLMs with reprompting architecture as a vital component toward realizing fully autonomous and intelligent industrial systems.




\bibliography{ifacconf}             
                                                   







\end{document}